\documentclass[twocolumn,showpacs,preprintnumbers,superscriptaddress,amsmath,amssymb]{revtex4-1}

\usepackage{epsfig}
\usepackage{graphicx}
\usepackage{dcolumn}
\usepackage{bm}
\usepackage{times}
\usepackage{xcolor}

\begin{document}


\title{Dynamics of social contagions with memory of non-redundant information}

\author{Wei Wang}
\affiliation{Web Sciences Center, University of Electronic
Science and Technology of China, Chengdu 610054, China}

\author{Ming Tang} \email{tangminghuang521@hotmail.com}
\affiliation{Web Sciences Center, University of Electronic
Science and Technology of China, Chengdu 610054, China}
\affiliation{State key Laboratory of Networking and Switching
Technology, Beijing University of Posts and Telecommunications,
Beijing 100876, China}

\author{Hai-Feng Zhang}\email{haifengzhang1978@gmail.com}
\affiliation{School of Mathematical Science, Anhui University,
Hefei 230039, China}

\author{Ying-Cheng Lai}
\affiliation{School of Electrical, Computer and Energy Engineering,
Arizona State University, Tempe, Arizona 85287, USA}

\date{\today}

\begin{abstract}

A key ingredient in social contagion dynamics is reinforcement,
as adopting a certain social behavior requires verification
of its credibility and legitimacy. Memory of
non-redundant information plays an important role
in reinforcement, which so far has eluded theoretical
analysis. We first propose a general social contagion model with
reinforcement derived from non-redundant information memory.
Then, we develop a unified edge-based compartmental theory
to analyze this model, and a remarkable agreement with
numerics is obtained on some specific models.
Using a spreading threshold model as a specific example to
understand the memory effect, in which each individual adopts
a social behavior only when the cumulative pieces of
information that the individual received from his/her
neighbors exceeds an adoption threshold. Through analysis and numerical simulations,
we find that the memory characteristic
markedly affects the dynamics as quantified by the final adoption size.
Strikingly, we uncover a transition phenomenon in which the dependence of
the final adoption size on some key parameters, such as the transmission
probability, can change from being discontinuous to being continuous.
The transition can be triggered by proper parameters
and structural perturbations to the system, such as decreasing
individuals' adoption threshold, increasing initial seed size, or enhancing the
network heterogeneity.

\end{abstract}

\pacs{89.75.Hc, 87.19.X-, 87.23.Ge}
\maketitle

\section{Introduction} \label{sec:intro}

Due to technological advances social networks are playing an ever
increasing role in the modern society. In a social network, nodes are
individuals of the population while links represent the social
ties or relations among individuals~\cite{Castellano2009}.
In recent years, there is a growing interest in investigating the
phenomenon of {\em behavior spreading} on social networks, where the
behaviors range from adoption of an innovation~\cite{Young2011} and
healthy activities~\cite{Centola2011} to microfinance~\cite{Banerjee2013}.
This is essentially the problem of {\em social contagion}. Ample
experimental and theoretical results indicated that, unlike biological
contagions in which successive contacts result in contagion with
independent probabilities, in a social contagion the probability of
infection depends on previous contacts. This is equivalent to
social affirmation or reinforcement effect,
since multiple confirmation of the credibility and legitimacy of the
behavior is always sought~\cite{Centola2007,Dodds2004,Dodds2005,
Centola2010,Weiss2014}. For an individual, who had two
friends adopting a particular behavior before a given time
and whose third friend newly adopts the behavior,
whether he/she adopts this behavior will take the three friends into account.

An early mathematical model to describe the dynamics of social
contagions is the threshold model~\cite{Granovetter1973,Watts2002}
based on Markovian process without memory, in which adoption of
behaviors depends only on the states of the current active neighbors
(i.e., individuals who have adopted the behavior), and
an individual adopts a behavior only when the current number or
the fraction of his/her active neighbors is equal to or exceeds some
adoption threshold. Analytically, the fraction of individuals adopting the behavior eventually,
can be predicted using the percolation theory~\cite{Watts2002} for
situations where the initial seed size is vanishingly small. One
result is that, for a fixed threshold, as the mean degree is increased,
the final size tends to grow continuously and then decrease
discontinuously. As the degree distribution becomes more heterogeneous,
the network is less vulnerable to social contagions, in sharp
contrast to the dynamics of epidemic spreading~\cite{Newman2002,
Pastor-Satorras2001,Boguna2013,Castellano2010}. Previous research
also revealed that, within the threshold model, factors such as
the initial seed size~\cite{Gleeson2007}, clustering
coefficient~\cite{Whitney2010}, community
structure~\cite{Gleeson2008,Nematzadeh2014},
multiplexity~\cite{Yagan2013,Brummitt2012,Lee2014}, and
temporal networks~\cite{Takaguchi2013,Karimi2013} all
can affect the social contagion process.

In real situations of social contagions, memory typically
plays an important role in the adoption and reinforcement of
behaviors, which includes full~\cite{Centola2011} or partial~\cite{Dodds2004} memory of the
cumulative behavioral information (behavioral
information can be referred as information for short) that individuals received
from their neighbors. This memory effect makes the
dynamics of social contagions have non-Markovian characteristic. To account
for the memory effect, sophisticated non-Markovian models
were proposed~\cite{Lv2011,Centola2011,Dodds2004,Dodds2005,Chung2014,Aral2012,Banerjee2013}.
In some models, it was predicted that the final adoption size will grow
discontinuously~\cite{Dodds2004,Dodds2005,Chung2014},
if the adoption probability for an individual who receives more than
one piece of information is two times larger than the adoption
probability for individuals getting only one piece of information. In general, the
memory of cumulative information
about the particular social behavior can come
from \emph{redundant}~\cite{Dodds2004,Dodds2005} or \emph{non-redundant}~\cite{Watts2002} information
transmission, where the former allows a pair
of individuals to transmit information successfully more
than once but for the latter, repetitive transmission
is forbidden. In some social contagion processes such as
risk migration and use of unproven technologies~\cite{Centola2007},
transmitting redundant information between the same pair of
individuals is unnecessary, since each neighbor can guarantee the
credibility and legitimacy of the behavior but only to certain
extent~\cite{Centola2011}. However, such non-redundant
information transmission characteristic of social reinforcement
have essentially been neglected in previous studies~\cite{Dodds2004,Dodds2005}.

A systematic study to understand the effects of non-redundant
information memory on social contagion dynamics is thus called for. A
general model needs to include different situations of behavior adoption
such as the dependence of the adoption probability on non-redundant
information~\cite{Centola2011} or even on the structure diversity of
such information~\cite{Ugander2012}. Due to the non-Markovian nature of
the memory characteristic, to develop a general theory
is challenging. Some approximate approaches were devised such as those based
on mean field analysis~\cite{Dodds2004}, percolation theory~\cite{Chung2014},
and renewal process~\cite{Mieghem2013,Cator2013}. Since the non-Markovian
property induces strong dynamical correlations between any two connected
individuals, analytic predictions from these approaches tend to deviate
significantly from results from direct numerical simulations, especially
when the underlying network is strongly structurally
heterogeneous~\cite{Pastor-Satorras2014}.

In this paper, we articulate a general social contagion model with social
reinforcement derived from memory of non-redundant information to address
the general question of how behaviors spread on networks in a more systematic
and complete way. In order to understand, quantitatively, the effects of
this kind of memory characteristic on social contagion
dynamics, we develop a unified edge-based compartmental theory. We base
our study on the spreading threshold model, focusing on the final behavior
adoption size and its dependence on the transmission probability under
different dynamical and topological parameter settings. We find that
the memory characteristic generally have a strong effect on
the final adoption size. Surprisingly, we uncover a crossover between
discontinuous and continuous variations in the final adoption size.
More specifically, the crossover phenomenon can be induced by decreasing
individuals' adoption threshold, increasing the initial
seed size or enhancing the structural heterogeneity of the
network. Our theoretical predictions
agree well with results from numerical simulations. We further
generalize our theory to treat distinct social contagion models and
network structures.

In Sec.~\ref{sec:model}, we describe our general social contagion model
with reinforcement derived from memory of non-redundant information
on complex networks. In Sec.~\ref{sec:theory}, we detail our edge-based
compartmental theory and analysis. In Sec.~\ref{sec:numerics}, we present
results from extensive numerical computations to validate our theory.
In Sec.~\ref{sec:other_models}, we extend our theoretical framework to
analyze alternative social contagion models, demonstrating the generality
of our theory. In Sec.~\ref{sec:conclusion}, we present conclusions
and discussions.

\section{A General Social Contagion Model} \label{sec:model}

Our goal is to construct a general stochastic model for social contagion dynamics,
taking into account social reinforcement through
\emph{non-redundant information memory} characteristic.
In this model, information refers to the behavioral information.  
The non-redundant information memory has two features:
(1) non-redundant information transmission, i.e., repetitive information transmission 
on every edge is forbidden, and thus also can be called as \emph{single-transmission};
(2) every individual can remember the cumulative pieces of non-redundant information
that the individual received from his/her neighbors, which makes the contagion processes be non-Markovian.

Concretely, we consider a configuration network model~\cite{Catanzaro2005} of
size $N$ and degree distribution $P(k)$, where nodes in the network
represent individuals. There is no degree-degree correlations
when the network is very large and sparse.
At any time, each individual can exist in one of
the three different states: susceptible, adopted, or recovered.
In the susceptible state, an individual does not adopt the social
behavior. In the adopted state, an individual adopts the behavior and
transmits the behavioral information to his/her neighbors. In the
recovered state, an individual loses interest in the behavior and
will not spread the  information further. This is thus a
susceptible-adopted-recovered (SAR) model. Although this proposed model
has similar state definitions with the epidemiology susceptible -
infected - recovered (SIR) model~\cite{Newman2002}, the non-markovian
characteristic is absent in the SIR model.

\begin{figure}
\begin{center}
\epsfig{file=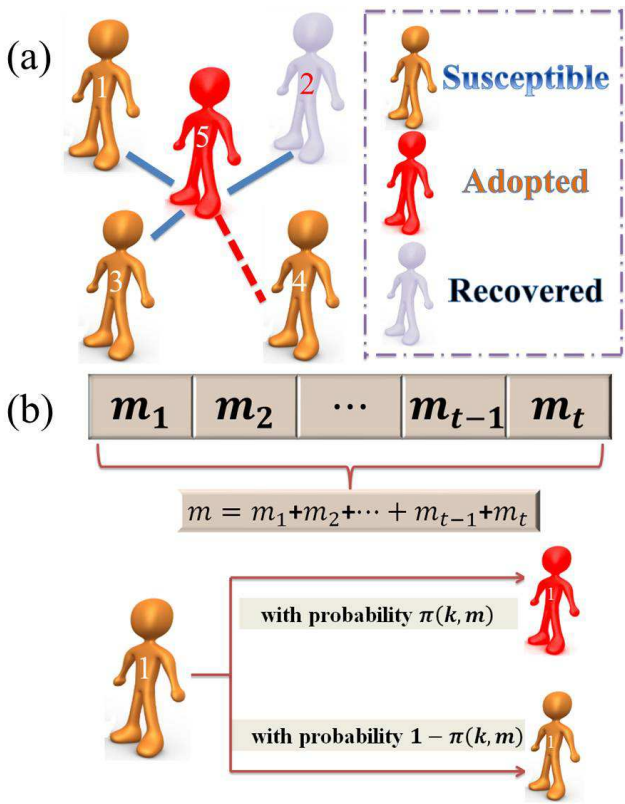,width=1\linewidth}
\caption{(Color online) {{\bf Illustration of the susceptible-adopted-recovered
(SAR) model on complex networks.} (a) At time $t$, the adopted individual $5$
tries to transmit the behavioral information (or simply information
for short) to each susceptible neighbor individual independently with
probability $\lambda$. Note that individual $5$ can not transmit the information
to individual $4$, since he/she has transmitted the information to individual
$4$ successfully before time $t$. That is to say, the susceptible individual
can only get the non-redundant information from his/her neighbors.
The solid blue lines denote that the information
has not transmitted through them successfully, and the red dished line denotes that the
information has transmitted through it previously.
(b) Assuming that individual $1$ has
received a new piece of information at time $t$, whether individual $1$ adopts
the behavior is determined by the $m$ cumulative pieces of information he/she ever
received from neighbors. The value of $m$ can be expressed as $m=\sum_{d=1}^tm_d$, where
$m_d$ is the pieces of information that individual $1$ received at time $d$.
In such a situation, individual $1$ has to remember the pieces of non-redundant
information he/she received from neighbors before time $t$. Thus, the so called
non-redundant information memory is induced.
Individual $1$ becomes adopted with probability $\pi(k,m)$, where $k$
is the degree of individual $1$; otherwise, individual $1$ remains in the susceptible
state. }}
\label{illu}
\end{center}
\end{figure}

To initiate a social contagion, a fraction $\rho_0$ of individuals
are uniformly randomly chosen to be in the adopted state and the remaining
majority of the individuals are in the susceptible state. At each time
step, behavioral information propagates from each adopted individual to
each neighbor independently with transmission probability $\lambda$, a key
parameter of the underlying dynamical process. We
assume an edge that has transmitted the information successfully
will never transmit the same information again, i.e., non-redundant 
information transmission. The non-redundant information
spreading process is illustrated schematically in Fig.~\ref{illu}(a). 
Based on this setting, we introduce the memory effect of non-redundant information
in social reinforcement. In particular,
assume that a susceptible individual $u$ of degree $k$ already has $m-1$
pieces of information from distinct neighbors. Once $u$ is successfully informed of the social
behavior by one of his/her adopted neighbors, denoted as $v$, the
cumulative number of pieces of information that $u$ has will increase by $1$.
With the $m$ cumulative pieces of information up to now
(i.e., after exposing to $m$ pieces of non-redundant information), the probability that the individual will be in the adopted
state is $\pi(k,m)$. Note that $u$ may subsequently get more than one
pieces of information successfully in this time step, thus, he/she will try to adopt
the behavior when he/she gets every new piece of information. In this case, if $u$
gets the $(n+1)$th new information in this step, he/she will adopt the behavior
with probability $\pi(k,m+n)$. An illustration
of the behavior adoption process is presented in Fig.~\ref{illu}(b). 
Since $\pi(k,m) < 1$ in general, multiple information transmission
is necessary for $u$ to move into the adopted
state, thereby incorporating the memory characteristic into the model. Generally, $\pi(k,m)$
is a monotonically increasing function of $m$ for any given degree $k$,
which characterizes the reinforcement effect through non-redundant information memory.
If $\pi(k,m)$ is a constant, no such reinforcement effect exists.
In this case, if the adopted state is regarded as the infected
state in epidemiology, our model
will reduce to the standard SIR (susceptible -
infected - recovered) model~\cite{Newman2002}, where
social reinforcement effect and non-Markovian properties are not present -
a key difference between biological and social contagions.
Empirical researches indicate
that the adopted individuals may lose interest in the behavior~\cite{Karsai2014},
which is also concerned in the binary social dynamics~\cite{Dodds2013,Harris2013}.
At the same time step, we thus assume that each adopted individual loses
interest in transmitting the behavioral information and becomes recovered
with probability $\gamma$. The spreading dynamics terminates once all
adopted individuals have become recovered.

By setting the parameters, our stochastic model
can generate either Markovian or
non-Markovian processes, thereby including a number of existing models
on social contagions as different limiting cases. For example, if
$\pi(k,m)$ is a Heaviside step function (i.e., if $m$ is less than
the adoption threshold $T_u$, then $\pi$ is zero; otherwise, $\pi$ is
unity), and setting $\lambda=1.0$ and $\gamma=0.0$, we obtain the Watts
threshold model~\cite{Watts2002}. Once the thresholds of individuals and network
topology are fixed, the cascade process in the Watts threshold model
will be deterministic, which is a trivial case of Markovian process.
In addition, by choosing the dynamical parameters properly, we can map our model into
some of the existing non-Markovian models. For instance, fixing
$\lambda=1.0$ and letting $\pi(k,m)$ be a function of exactly one of the two 
quantities (i.e., adopted and susceptible individuals), we recover the synergy spreading
model~\cite{Francisco2011}. Similarly, if we allow $\pi(k,m)$ to be a
linear~\cite{Chung2014} or exponential~\cite{Lv2011} function of $m$ and $\gamma=1.0$,
we can obtain distinct types of non-Markovian dynamics.
Differing from the models in Refs.~\cite{Chung2014,Lv2011} in which each adopted individual
only gets one chance to transmit the behavioral information to every neighbor, in our model an
adopted individual can try to transmit the information many times until
he/she becomes recovered state or transmits the information successfully.

In our study, we concentrate on the so-called spreading threshold model
before turning to more generalized social contagion models. In the
spreading threshold model, an individual $u$ adopts the behavior only
when the number of pieces of non-redundant  information that
$u$ possessed exceeds the adoption threshold $T_u$. This means that
the adoption probability $\pi(k,m)$ is a Heaviside step function, which has
the same form as in the Watts threshold model~\cite{Watts2002}. There are,
however, key differences between the two types of threshold models.
Firstly, differing from the Watts model in which the adoption threshold
is the corresponding fraction of neighboring nodes, the
adoption threshold in our model is expressed in terms of the absolute
number of neighboring nodes, as in bootstrap percolation~\cite{Baxter2010}
and self-organized criticality models~\cite{Zhang1989}. Secondly, in the
Watts model, each individual can obtain information about
the states of all its neighbors ``instantaneously'' at each time step,
but in our model individuals are able to know the neighboring states
only through transmission of the  information. Thirdly, in
the Watts model an individual is permanently interested in the behavior
even after its adoption, while we assume more realistically that
individuals having adopted certain behavior may lose interest in it
and never spread the corresponding information, which is quantified by
the abandon probability $\gamma$. Note that
if the threshold of Watts model is expressed as the absolute number of
neighbors who have adopted the behavior, there will
only exist the second and third differences. These three differences are
consequences of introducing the non-redundant information memory characteristic
into our model, better capturing the essential dynamics of social
contagions in the real world.

\section{Theory} \label{sec:theory}

We first develop a unified edge-based compartmental theory to analyze our
general social contagion model with reinforcement mechanism based on
non-redundant information memory characteristic. We then systematically
investigate how the memory affects the social contagion process in a
specific model, the spreading threshold model. In this theory,
we assume that the networks have large network sizes, sparse edges,
and no degree-degree correlations, and the contagion dynamics evolves continuously.
Mathematically, a contagion process can be described by three variables:
$S(t)$, $A(t)$ and $R(t)$, which are the densities of the susceptible,
adopted, and recovered individuals at time $t$, respectively.
The states of all individuals remain unchanged when $t\to\infty$,
and $R(\infty)$ is the final fraction of individuals that
have adopted the social behavior.

\subsection{General theoretical framework} \label{subsec:theory_gen}

Due to the non-redundant information memory characteristic, in a social contagion
process there are strong dynamical correlations between the states of the
adjacent nodes, making existing theoretical methods such as the
mean-field theory~\cite{Dodds2004}, percolation theory~\cite{Gleeson2007},
and renewal process~\cite{Cator2013} inapplicable, especially for networks
that are strongly structurally heterogeneous. Using insights from
Refs.~\cite{Miller2011,Miller2013,Wang2014A,Valdez2012}, we develop
an edge-based compartmental theory to analyze social contagion dynamics
in the presence of strong nodal state correlations.

Let $\theta(t)$ be the probability that individual $v$ has not transmitted
the  information to individual $u$ along a randomly chosen edge
by time $t$. In the spirit of the cavity theory~\cite{Karrer2010,Miller2013},
we disallow individual $u$ to transmit any  information to its
neighbors but $u$ can receive such information from its neighbors -
$u$ is in a cavity state. Initially, a fraction of $\rho_0$ individuals
is in the adopted state, and none of them transmits the
information to its neighbors, so $\theta(0)=1$ for all edges.
For simplicity in theory, we assume that the probability of not transmitting
the information is identical for all edges, and dynamical correlations
doesn't exist among neighbors of an individual.
At time $t$, a uniformly randomly chosen
individual $u$ of degree $k$ in the cavity state has been
exposed to $m$ pieces of non-redundant information (i.e.,
$u$ has received the information from distinct neighbors $m$
times) with the probability
\begin{equation} \label{active_a_k}
\phi_m(k,\theta(t))=(1-\rho_0)\binom{k}{m}\theta(t)^{k-m}[1-\theta(t)]^m,
\end{equation}
where the factor $(1-\rho_0)$ is the fraction of
susceptible nodes initial. By time $t$, the susceptible individual $u$ has received the
 information from $m$ different neighbors. The probability that
$u$ has not adopted the behavior for time of receiving  information
less than $m$ is $\Pi_{j=0}^m[1-\pi(k,j)]$. Combining this factor and summing
over all possible values of $m$, we obtain the probability that the individual
$u$ is still in the susceptible state at time $t$ as
\begin{equation} \label{S_K_T}
s(k,t)=\sum_{m=0}^{k}\phi_m(k,t)\Pi_{j=0}^m[1-\pi(k,j)].
\end{equation}
Taking into account different degrees in the network, we obtain the fraction
of susceptible individuals (i.e., the probability of a randomly chosen
individual is in the susceptible state) at time $t$ as
\begin{equation} \label{s_t}
S(t)=\sum_{k=0}^{\infty}P(k)s(k,t).
\end{equation}
Analogously, the fraction of individuals with $m$ pieces of
information at time $t$ is
\begin{equation} \label{Phi_n_T}
\Phi(m,t)=\sum_{k=0}^{\infty}P(k)\phi_m(k,\theta(t)).
\end{equation}
A neighbor of individual $u$ may be in one of susceptible, adopted, or
recovered states. We can thus further express $\theta(t)$ as
\begin{equation} \label{theta}
\theta(t)=\xi_S(t)+\xi_A(t)+\xi_R(t),
\end{equation}
where $\xi_S(t)$ [$\xi_A(t)$ or $\xi_R(t)$] is the probability that a
neighbor of the individual $u$ in the cavity state is in the susceptible
(adopted or recovered) state and has not
transmitted the  information to individual $u$ through an edge by time $t$.
Note that the three quantities are unknown, which are to be solved.

If a neighboring individual $v$ of $u$ is initially in the
susceptible state with probability $1-\rho_0$, it cannot
transmit the information to $u$. Individual $v$ can get the
 information from its other neighbors, since $u$ is in a cavity
state. At time $t$, the probability
that individual $v$ has received $m$ pieces of non-redundant information is
\begin{equation} \label{tao_S}
\tau_m(k^{\prime},\theta(t))=\binom{k^{\prime}-1}{m}\theta(t)^{k^{\prime}-m-1}
[1-\theta(t)]^m,
\end{equation}
where $k^{\prime}$ is the degree of $v$. Similar to Eq.~(\ref{S_K_T}),
individual $v$ will still stay in the susceptible state at time $t$ with
the probability
\begin{equation} \label{neighbour_S}
\Theta(k^{\prime},\theta(t))=\sum_{m=0}^{k^{\prime}-1}\tau_m(k^{\prime},t)
\Pi_{j=0}^{m}[1-\pi(k^{\prime},j)].
\end{equation}
For uncorrelated networks, the probability that one edge from individual $u$
connects with an individual with degree $k^{\prime}$ is
$k^{\prime}P(k)/\langle k\rangle$, where $\langle k\rangle$ is the mean degree
of the network. Summing over all possible $k^{\prime}$, we obtain the
probability that $u$ connects to a susceptible individual by time $t$ as
\begin{equation} \label{xi_S}
\xi_S(t)=(1-\rho_0)\frac{\sum_{k^{\prime}}k^{\prime}P(k)
\Theta(k^{\prime},\theta(t))}{\langle k\rangle}.
\end{equation}
The information spreading process as described in Sec.~\ref{sec:model} suggests
that two events need to occur in order for the growth of $\xi_R$:
(1) with probability $1-\lambda$ an adopted neighbor has not transmitted
the  information to $u$ via their connection and (2) with
probability $\gamma$ the adopted neighbor has been recovered.
Taking these into consideration, we get
\begin{equation} \label{xi_R}
\frac{d\xi_R(t)}{dt}=\gamma(1-\lambda)\xi_A(t).
\end{equation}

At time $t$, the rate of change in the probability that a random edge
has not transmitted the information is equal to the rate at which
the adopted neighbors transmit the information to their
susceptible neighboring individuals through edges. Thus, we get  
\begin{equation} \label{d_theta}
\frac{d\theta(t)}{dt}=-\lambda\xi_A(t).
\end{equation}
Combining Eqs.~(\ref{xi_R}) and (\ref{d_theta}), we obtain
\begin{equation} \label{xi_R_2}
\xi_R(t)=\frac{\gamma[1-\theta(t)](1-\lambda)}{\lambda}.
\end{equation}
Substituting Eqs.~(\ref{xi_S}) and (\ref{xi_R_2}) into Eq.~(\ref{theta}),
we get an expression for $\xi_A(t)$ in terms of $\theta(t)$. Doing so, we
can rewrite Eq.~(\ref{d_theta}) as
\begin{equation} \label{d_theta_2}
\begin{split}
\frac{d\theta(t)}{dt}&=-\lambda[\theta(t)-(1-\rho_0)
\frac{\sum_{k^{\prime}} k^{\prime}P(k^{\prime})\Theta(k^{\prime},\theta(t))}
{\langle k\rangle}] \\
&+\gamma[1-\theta(t)](1-\lambda).
\end{split}
\end{equation}
Note that the rate $dA(t)/dt$ is equal to the rate at which $S(t)$
decreases, because all the individuals moving out of the susceptible state
must move into the adopted state minus the rate at which adopted individuals
become recovered. We have
\begin{equation} \label{rho_t}
\frac{dA(t)}{dt} = -\frac{dS(t)}{dt} - \gamma A(t)
\end{equation}
and
\begin{equation} \label{r_T}
\frac{dR(t)}{dt}=\gamma A(t).
\end{equation}
Equations~(\ref{active_a_k})-(\ref{s_t}) and (\ref{d_theta_2})-(\ref{r_T})
give us a complete and general description of social contagion dynamics,
from which the density for each type of individual in each state at arbitrary
time step can be calculated.

Say we are interested in the steady state of the social contagion dynamics.
Setting the right side of Eq.~(\ref{d_theta_2}) to be zero, we get
\begin{equation} \label{stady_theta}
\begin{split}
\theta(\infty)&=(1-\rho_0)\frac{\sum_{k^{\prime}} k^{\prime}P(k^{\prime})
\Theta(k^{\prime},\theta(\infty))}{\langle k\rangle}\\
&+\frac{\gamma[1-\theta(\infty)](1-\lambda)}{\lambda},
\end{split}
\end{equation}
where $\Theta(k^{\prime},\theta(\infty))$ is
a nonlinear function of $\theta(\infty)$.
Note that $\theta(t)$ decreases with $t$, as the individuals in the adopted
state persistently transmit the information to their neighbors.
Thus in simulations, only the maximum
value of the stable fixed point (if there exist more than one stable fixed points)
of Eq.~(\ref{stady_theta}) is physically meaningful. Substituting this value into
Eqs.~(\ref{active_a_k})-(\ref{s_t}), we can obtain the value of the susceptible
density $S(\infty)$ and the final behavior adoption size $R(\infty)$.

As in epidemic spreading, the condition under which outbreak of
behavior adoption occurs is of interest. Similar to analyzing epidemic spreading,
we can obtain the critical condition by determining when a nontrivial solution
of Eq.~(\ref{stady_theta}) appears, which corresponds to the point at which
the equation
\begin{eqnarray}
\nonumber
g[\theta(\infty),\rho_0,T,\gamma,\lambda]
&= (1-\rho_0)\frac{\sum_{k^{\prime}}
k^{\prime}P(k^{\prime})\Theta(k^{\prime},\theta(\infty))}{\langle k\rangle}\\ \nonumber
&+\frac{\gamma[1-\theta(\infty)](1-\lambda)}{\lambda}
-\theta(\infty)
\end{eqnarray}
is tangent to horizontal axis at the critical value of $\theta_c(\infty)$.
The value of $\theta_c(\infty)$ denotes the critical probability that the
information is not transmitted to $u$ via an edge at the critical transmission
probability when $t\to\infty$. This way we obtain the critical condition
of the general social contagion model as
\begin{equation} \label{First_Order}
\frac{dg}{d\theta(\infty)}|_{\theta_c(\infty)}=0.
\end{equation}
From Eq.~(\ref{First_Order}), we can calculate the critical transmission
probability:
\begin{equation} \label{First_Order_Exp}
\lambda_c=\frac{\gamma}{\Delta+\gamma-1},
\end{equation}
where
\begin{displaymath}
\Delta=(1-\rho_0)\frac{\sum_{k^{\prime}} k^{\prime}P(k^{\prime})
\frac{d\Theta(k^{\prime},\theta(\infty))}{d\theta(\infty)}|_{\theta_c(\infty)}}
{\langle k\rangle}.
\end{displaymath}
From Eqs.~(\ref{tao_S})-(\ref{neighbour_S}), we obtain
the expression of $\frac{d\Theta(k^{\prime},\theta(\infty))}{d\theta(\infty)}$ as
\begin{equation} \label{d_theta_Fianl}
\begin{split}
\frac{d\Theta(k^{\prime},\theta(\infty))}{d\theta(\infty)}
&=\sum_{m=0}^{k^{\prime}-1}\binom{k^{\prime}-1}{m}\\
&\times\{
(k^{\prime}-m-1)\theta(\infty)^{k^{\prime}-m-2}[1-\theta(\infty)]^m\\
&-m\theta(\infty)^{k^{\prime}-m-1}[1-\theta(\infty)]^{m-1}\}\\
&\times\Pi_{j=0}^{m}[1-\pi(k^{\prime},j)].
\end{split}
\end{equation}
Numerically solving Eqs.~(\ref{stady_theta}) and
(\ref{First_Order_Exp})-(\ref{d_theta_Fianl}), we can get the critical value
of the transmission probability $\lambda_c$ for any given adoption function
$\pi(k,m)$. We see that $\lambda_c$ is correlated with the dynamical
parameters such as the adoption probability $\pi(k,m)$, the initial seed
size $\rho_0$ and the abandon probability $\gamma$, as well as the
topological parameters of the network [e.g., the degree distribution
$P(k)$ and the mean degree $\langle k\rangle$].

\subsection{Spreading threshold model}
We now apply the general theoretical framework developed in
Sec.~\ref{subsec:theory_gen} to analyzing a specific class of social
contagion model - spreading threshold model. In this model, the adoption
function $\pi(k,m)$ is a Heaviside step function:
\begin{equation} \label{threshold_model}
\pi(k,m)= \begin{cases}
1, ~m\geq T_k,\\
0, ~m<T_k,
\end{cases}
\end{equation}
where $T_k$ is the adoption threshold of individuals of degree $k$.
Here the adoption probability $\pi(k,m)$ is only a function of $k$ and $m$.
Further investigations on general model, incorporating individuals' inherent
characters such as age and habit, are called
for. Utilizing Eq.~(\ref{threshold_model}),
we can write Eqs.~(\ref{S_K_T}) and (\ref{neighbour_S}) as
\begin{equation} \label{S_K_T_SPE}
s(k,t)=\sum_{m=0}^{T_k-1}\phi_m(k,\theta(t))
\end{equation}
and
\begin{equation} \label{neighbour_S_SPE}
\Theta(k^{\prime},\theta(t))=\sum_{m=0}^{T_{k^{\prime}}-1}\tau_m(k^{\prime},\theta(t)),
\end{equation}
respectively. Similarly, Eq.~(\ref{rho_t}) becomes
\begin{equation} \label{rho_t_SPE}
\frac{dA(t)}{dt}=\lambda\xi_A\psi(t)-\gamma A(t),
\end{equation}
where
\begin{equation} \label{psi}
\begin{split}
\psi(t)&=(1-\rho_0)\sum_{k=0}^{\infty}P(k)\sum_{m=0}^{T_k-1}\binom{k}{m}\\
&\times\{(k-m)\theta(t)^{k-m-1}[1-\theta(t)]^m\\
&-m\theta(t)^{k-m}[1-\theta(t)]^{m-1}\}.
\end{split}
\end{equation}
The critical condition can be determined using Eq.~(\ref{First_Order}).
For the simple case where the fraction of randomly chosen initial seeds
is vanishingly small (i.e., $\rho_0 \rightarrow 0$) and all individuals
with different degrees
have the same adoption threshold $T$, Eq.~(\ref{stady_theta}) has one
trivial solution: $\theta(\infty)=1$. At the critical point, the function
$g[\theta(\infty),\rho_0,T,\gamma,\lambda]$ is tangent to horizontal axis
at $\theta(\infty)=1$. For $T=1$, using
Eqs.~(\ref{stady_theta})-(\ref{d_theta_Fianl}), we obtain the
continuous critical transmission probability as
\begin{equation} \label{Second_Order}
\lambda_c^{II}=\frac{\gamma\langle k\rangle}{\langle k^2\rangle -
2\langle k\rangle+\gamma\langle k\rangle},
\end{equation}
which has the same form as the epidemic outbreak threshold.
However, for $T>1$, the function $g[\theta(\infty),\rho_0,T,\gamma,\lambda]$
can never be tangent to horizontal axis, suggesting
that a vanishingly small fraction of initial seeds cannot cause a finite
fraction of the individuals to adopt the behavior.

\begin{figure}
\begin{center}
\epsfig{file=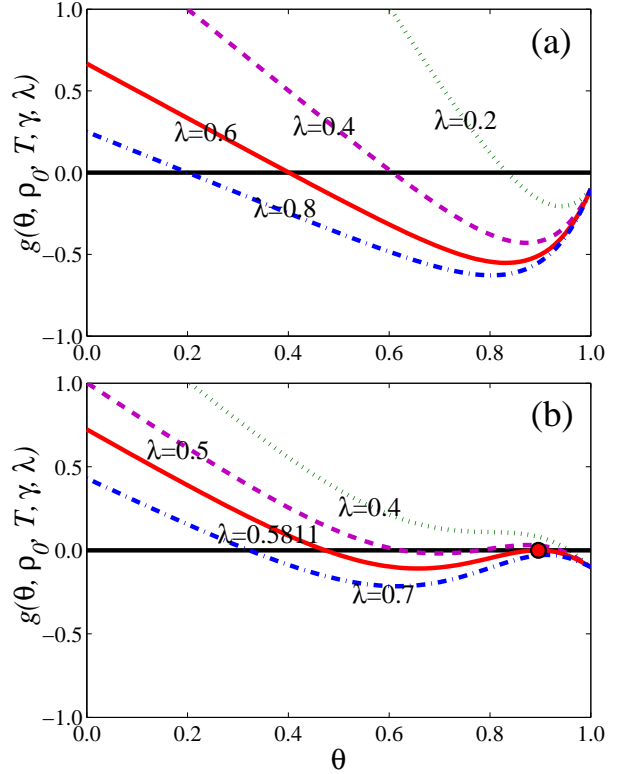,width=1\linewidth}
\caption{(Color online) {\bf Illustration of graphical solutions of
Eq.~(\ref{stady_theta}).} For ER random networks, (a) continuously
increasing behavior of $R(\infty)$ with $\lambda$ for $T=1$,
(b) discontinuous change in $R(\infty)$ for $T=3$. The
black solid lines are the horizontal axis
and the red dots denote the tangent points. Other parameters are
$\rho_0=0.1$ and $\gamma=1.0$.}
\label{fig1}
\end{center}
\end{figure}

Now suppose that $\rho_0$ is not vanishingly small so that $\theta(\infty)=1$
is no longer a solution of Eq.~(\ref{stady_theta}). In this case, regardless
of the values of other parameters, a finite fraction of individuals will
adopt the behavior. It is thus reasonable to focus on how
non-redundant information memory
characteristic affects the dependence of the final
behavioral adoption size $R(\infty)$ on the transmission probability $\lambda$,
which can be obtained from the roots of Eq.~(\ref{stady_theta}). We are
particularly interested in finding out whether the dependence is continuous
or discontinuous. Note that the number of roots Eq.~(\ref{stady_theta})
is odd (including multiplicity) for any parameters, because of the
$g[\theta(\infty),\rho_0,T,\gamma,\lambda]<0$
for $\theta(\infty)=1$ and
$g[\theta(\infty),\rho_0,T,\gamma,\lambda]>0$
for $\theta(\infty)=0$. As shown in Figs.~\ref{fig1}(a) and \ref{fig1}(b),
numerical calculations indicate that the number of roots is either 1 or 3.
When we fix all the parameters except $\lambda$, if Eq.~(\ref{stady_theta})
has only one root for different values of $\lambda$ [Fig.~\ref{fig1}(a)],
$R(\infty)$ will increase continuously with $\lambda$.
If the number of roots of Eq.~(\ref{stady_theta}) depends on $\lambda$, as
shown in Fig.~\ref{fig1}(b), there will be three roots (fixed points),
which means a saddle-node bifurcation occurs~\cite{Strogatz2005}. The
bifurcation analysis of Eq.~(\ref{stady_theta})
reveals that the system undergoes a cusp catastrophe: Varying $\lambda$,
one finds that the physically meaningful stable solution of $\theta(\infty)$
will suddenly jump to an alternate outcome. In this case, a
discontinuous growth pattern of $R(\infty)$ with $\lambda$
emerges, and the critical transmission probability $\lambda_c^I$ at which
the discontinuity occurs can be obtained by solving Eqs.~(\ref{stady_theta})
and (\ref{First_Order_Exp})-(\ref{d_theta_Fianl}).

The discontinuous behavior in $R(\infty)$ versus $\lambda$
can be understood by using a specific example, e.g., by setting $T = 3$.
As shown in Fig.~\ref{fig1}(b), for different values of $\lambda$, the
function $g[\theta(\infty),\rho_0,T,\gamma,\lambda]$ is tangent to
horizontal axis at $\lambda_c^I \approx 0.5811$. For $\lambda<\lambda_c^I$,
if Eq.~(\ref{stady_theta}) has 3 fixed points then the solution will be
given by the largest one (since only this value can be achieved physically).
Otherwise, the solution is the only fixed point. For $\lambda=\lambda_c^I$,
the solution is given by the tangent point. For $\lambda>\lambda_c^I$, the
only fixed point is the solution of Eq.~(\ref{stady_theta}). In this case,
the solution of Eq.~(\ref{stady_theta}) changes abruptly to a small value
from a relatively large value at $\lambda=\lambda_c^I$, leading to a
discontinuous change in $R(\infty)$.

Finally, to determine the critical system parameter value of
$\theta_s(\infty)$, across which the dependence of $R(\infty)$ on
$\lambda$ changes from being continuous (discontinuous) to discontinuous
(continuous), we can numerically solve Eqs.~(\ref{stady_theta}) and
(\ref{First_Order}) together with the condition
\begin{equation} \label{First_Order_Condition}
\frac{d^2g[\theta(\infty),\rho_0,T,\gamma,\lambda]}{d\theta^2(\infty)}=0.
\end{equation}
From Eq.~(\ref{First_Order_Condition}), we obtain
\begin{equation} \label{rho_S}
\rho_0^s=\frac{1}{\digamma},
\end{equation}
where $\digamma=\sum_{k^{\prime}}k^{\prime}P(k^{\prime})
\frac{d\Theta^2(k^{\prime},\theta(\infty))}{d\theta^2(\infty)}$.
Using Eqs.~(\ref{tao_S}) and (\ref{neighbour_S_SPE}), we get
\begin{equation} \label{d_theta_Fianl_2}
\begin{split}
\frac{d\Theta^2(k^{\prime},\theta(\infty))}{d\theta^2(\infty)}
&=\sum_{m=0}^{T_{k^{\prime}}-1}\binom{k^{\prime}-1}{m}\\
&\times\{
(k^{\prime}-m-1)[(k^{\prime}-m-2)\theta(\infty)^{k^{\prime}-m-3}\\
&\times(1-\theta(\infty))^m-m\theta(\infty)^{k^{\prime}-m-2}
(1-\theta(\infty))^{m-1}]\\
&-m[(k^{\prime}-m-1)\theta(\infty)^{k^{\prime}-m-2}(1-\theta(\infty))^{m-1}\\
&-(m-1)\theta(\infty)^{k^{\prime}-m-1}(1-\theta(\infty))^{m-2}]\}.
\end{split}
\end{equation}
Combining Eqs.~(\ref{stady_theta}), (\ref{First_Order}) and
(\ref{First_Order_Condition}), we get the value of $\theta_s(\infty)$.
For fixed $T$ and $P(k)$, the critical values of other system parameters
e.g., $\lambda_c^s$ and $\rho_0^s$, can then be determined.

\section{Numerical verification} \label{sec:numerics}

We perform extensive simulations on the spreading threshold model.
In our simulations, we use Erd\H{o}s-R\'{e}nyi (ER) network model~\cite{Erdos1959} and
configuration network model with power-law degree distribution~\cite{Catanzaro2005},
where the network size and mean degree are $N=10^4$ and
$\langle k\rangle=10$, respectively. At least $2\times10^3$
independent dynamical realizations on a fixed network are used to calculate
the pertinent average values, which are further averaged over $100$ network
realizations. We separately discuss the
effects of dynamical and topological parameters.

\subsection{Effects of dynamical parameters}
\label{subsec:dynamical_parameters}

\begin{figure}
\begin{center}
\epsfig{file=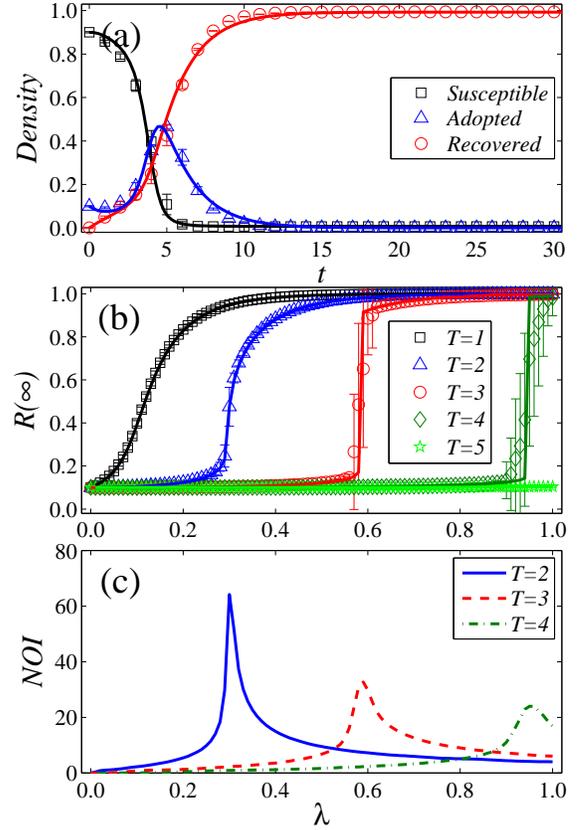,width=1\linewidth}
\caption{(Color online) {\bf Spreading threshold model on ER networks.}
(a) Average densities of susceptible, adopted, and recovered populations,
denoted by $S(t)$ (black squares), $A(t)$ (blue up triangles), and $R(t)$
(red circles), respectively, versus time. (b) Final behavior adoption
size $R(\infty)$ versus the transmission probability $\lambda$ for $T=1$
(black squares), $T=2$ (blue up triangles), $T=3$ (red circles), $T=4$
(dark green diamonds), and $T=5$ (light green stars) in the steady state.
(c) Simulation results of NOI (number of iterations) as a function of
$\lambda$ with $T=2$ (blue solid line), $T=3$ (red dashed line) and $T=4$
(dark dash dotted green
line). The error bars indicate the standard deviations. The lines in (a)
and (b) are theoretical predictions based on solutions of
Eqs.~(\ref{active_a_k})-(\ref{s_t}) and (\ref{d_theta_2})-(\ref{r_T}).
In (a), we set $\lambda=0.8$, $\rho_0=0.1$, $T=3$, and $\gamma=0.5$ (so
as to obtain longer evolution time), while in (b) and (c), we set
$\rho_0=0.1$ and $\gamma=1.0$.}
\label{fig2}
\end{center}
\end{figure}

To be illustrative, we use ER random
networks~\cite{Erdos1959}. We first calculate the time evolution of
the population densities for $\lambda=0.8$, $\rho_0=0.1$, $T=3$, and $\gamma=0.5$,
as shown in Fig.~\ref{fig2}(a), where we observe that the density of the susceptible (recovered) individuals
decreases (increases) with time, and reaches some final value for
large time. The density of the adopted individuals decreases
initially (due to the fact that the number of individuals who newly adopt
the behavior is less than that of individuals who become recovered), then
increases and reaches a maximum value at $t \approx 5$. These results agree
well with the predictions from our edge-based compartmental theory
[see lines in Fig.~\ref{fig2}(a)].

We next study the final behavior adoption size $R(\infty)$ as a function
of the transmission probability $\lambda$ for different values of the
adoption threshold $T$ at another value of $\gamma=1.0$. As shown in Fig.~\ref{fig2}(b), increasing $T$
impedes individuals from adopting the behavior, since a larger value of $T$
requires the individual to be exposed with more  information from
distinct neighbors to affirm the authority and legality of the behavior.
As a result, individuals hardly adopt the behavior when the adoption
threshold is relatively large (e.g., $T \geq 5$).
Lines from the theory in Fig.~\ref{fig2}(b) are very consistent with these simulation results.
Through the bifurcation analysis of Eq.~(\ref{stady_theta}), we note
that the adoption threshold affects strongly the manner by which $R(\infty)$
increases with $\lambda$ for $T \leq 4$. As shown in Fig.~\ref{fig2}(b),
for some small adoption threshold (e.g., $T=1$), $R(\infty)$ increases
continuously with $\lambda$. However, for a slightly larger adoption
threshold (i.e., $T \agt 1$), the $R(\infty)$ versus $\lambda$ pattern
becomes discontinuous, exhibiting an abrupt increase at some critical
value $\lambda_c^I$. The statistical errors are generally small except
for $\lambda$ close to $\lambda_c^I$ (for this reason and for figure
clarity the error bars will not be shown for subsequent figures).
The theoretical value of $\lambda_c^I$ can be calculated from
Eqs.~(\ref{stady_theta}) and (\ref{First_Order_Exp})-(\ref{d_theta_Fianl}).
The critical value can also be estimated by observing the number of
iterations~\cite{Parshani2011,Liu2012} (denoted as NOI, where only those
iterations at which at least one individual adopts the behavior are taken
into account). We observe that the NOI exhibits a maximum value at
$\lambda_c^I$, e.g., for $T=2$, $3$, and $4$, as shown in Fig.~\ref{fig2}(c).
Overall, there is a remarkable agreement between theory and numerics
in terms of the quantities $R(\infty)$ and $\lambda_c^I$.
Through extensive simulations and theoretical predictions,
we know the abandon probability doesn't qualitatively affect the growth patterns of $R(\infty)$,
so it is set as $\gamma=1.0$ in the rest of this paper.

\begin{figure}
\begin{center}
\epsfig{file=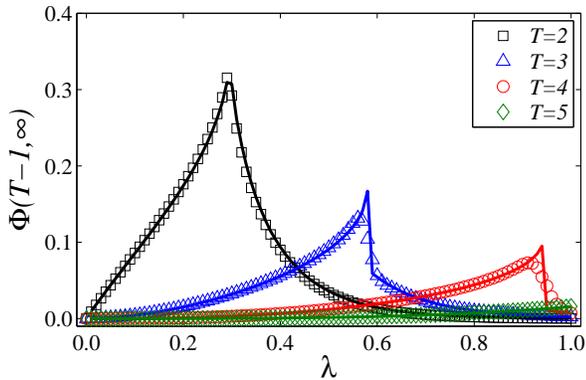,width=1\linewidth}
\caption{(Color online) \textbf{The final fraction of individuals in the subcritical
state on ER networks.} $\Phi(T-1,\infty)$ versus $\lambda$ for $T=2$
(black squares), $T=3$ (blue up triangles), $T=4$ (red circles) and $T=5$
(dark green diamonds). The lines are theoretical predictions based on solutions of
Eqs.~(\ref{active_a_k}), (\ref{Phi_n_T}) and (\ref{d_theta_2})-(\ref{r_T}). Other
parameters are $\gamma=1.0$ and $\rho_0=0.1$.}
\label{fig_subI}
\end{center}
\end{figure}

\begin{figure}
\begin{center}
\epsfig{file=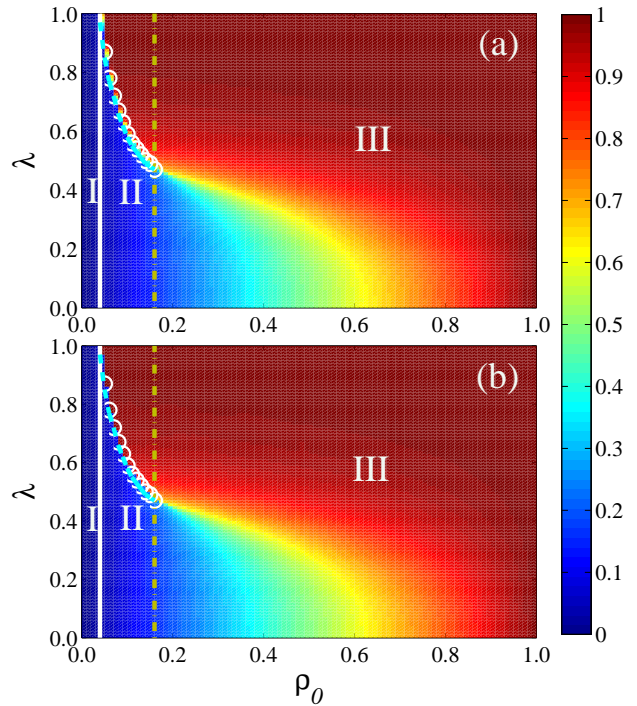,width=1\linewidth}
\caption{(Color online) {\bf Dependence of final behavior adoption size
on initial seed size and transmission probability.} For spreading threshold
model on ER random networks, color-coded values of $R(\infty)$ from numerical
simulations (a) and theoretical solutions (b) in the parameter plane
($\rho_0$,$\lambda$), where the theoretical values are from solutions of
Eqs.~(\ref{active_a_k})-(\ref{s_t}) and (\ref{d_theta_2})-(\ref{r_T}).
The numerically obtained critical values of the transmission probability,
$\lambda_c^I$ (white circles), are from the NOI method, and the corresponding theoretical
values blue dashed line) are from Eqs.~(\ref{stady_theta}) and
(\ref{First_Order_Exp})-(\ref{d_theta_Fianl}). In each subfigure,
three regions are shown: only a vanishingly small fraction of individuals can be exposed to
adopt the behavior in region I, in region II $R(\infty)$ grows discontinuously with $\lambda$
and a finite fraction of individuals adopt the behavior above $\lambda_c^I$,
and $R(\infty)$ grows continuously to a large value in region III. The vertical white
solid lines and dash dotted yellow lines
separate the plane into the three regions, which are predicted from
our edge-based compartmental theory. Other parameters are $\gamma=1.0$
and $T=3$.}
\label{fig3}
\end{center}
\end{figure}

It is useful to identify the key factors that affect the dependence of
$R(\infty)$ on $\lambda$. To obtain an intuitive understanding of the
phenomenon of abrupt increase in $R(\infty)$ as $\lambda$ passes through
a critical point, we focus on the individuals in the subcritical state.
An individual $u$ in such a state has received the  information
but has not yet adopted the behavior, and the number of information
pieces from distinct neighbors is precisely one less than the adoption
threshold. Say at the time $u$ has received information from his/her
neighbors except neighbor $v$. Now assume that $v$ has adopted the
social behavior and transmits the information to $u$ successfully
so as to cause $u$ to adopt the behavior. In turn, $u$ will transmit
the  information to his/her susceptible neighbors with
probability $\lambda$, which will cause some
subcritical state neighbors to adopt the
behavior accordingly, and so on, potentially leading to an avalanche
of behavior adoption. If the system has a relatively large number of
individuals in the subcritical state, a slight increase in the number
of individuals who adopt the behavior, e.g., by increasing the value
of $\lambda$ slightly, may cause a sudden and large number of such
subcritical state individuals with information pieces greater than
their threshold, leading to a discontinuous ``jump''
in the value of $R(\infty)$. The above intuitive understanding
is further proved by numerical simulations and theoretical predictions in Fig.~\ref{fig_subI}.
For $2\leq T<5$, the final fraction of individuals in subcritical state $\Phi(T-1,\infty)$
first increases with $\lambda$ below $\lambda_c^I$, $\Phi(T-1,\infty)$ reaches a maximum at the $\lambda_c^I$;
and a slight increment of $\lambda$ induces a finite fraction of $\Phi(T-1,\infty)$ to adopt the behavior simultaneously,
which leads to a discontinuous jump in the value of $R(\infty)$.
When this social reinforcement
effect is not present [e.g., $T = 1$ in Fig.~\ref{fig2}(b)], there are
essentially no individuals in the subcritical state. In this case,
$R(\infty)$ increases continuously with $\lambda$. We mention that
the mechanism underlying the discontinuous increase in $R(\infty)$
in our spreading threshold model is similar to that responsible for
phenomena such as explosive percolation~\cite{Achlioptas2010},
bootstrap percolation~\cite{Baxter2010}, \emph{k}-core
percolation~\cite{Dorogovtsev2006} and explosive
synchronization~\cite{Gomez-Gardenes2011}.

We further investigate the role of the initial seed size $\rho_0$ in
social contagion dynamics for relatively larger values of $T$, e.g., $T=3$.
As shown in Fig.~\ref{fig3}, we see that $R(\infty)$ increases with $\rho_0$,
since individuals in the network have more chances to be exposed to the
 information. Based on the values of $R(\infty)$, we can
divide the phase diagram into local ($\rho_0<0.04$) and global
($0.04\leq\rho_0\leq1$) behavior adoption regions, where in the former (i.e., region I),
only a vanishingly small fraction of individuals can be exposed to
adopt the behavior and, in the latter including regions II and III,
a finite fraction of individuals
adopt the behavior and a crossover phenomenon occurs
in the dependence of $R(\infty)$ on $\lambda$.
The crossover phenomenon means
that the dependence of $R(\infty)$ on $\lambda$ can change from
being discontinuous to being continuous. More specifically,
the saddle-node bifurcation of
Eq.~(\ref{stady_theta}) occurs for $0.04\leq\rho_0\leq0.15$ (region II
in Fig.~\ref{fig3}), thus $R(\infty)$ versus $\lambda$ is discontinuous;
$R(\infty)$ versus $\lambda$ is continuous for $0.15<\rho_0\leq1$ (region III in Fig.~\ref{fig3}),
as the saddle-node bifurcation disappears.
The crossover phenomenon originates from the fact that
the number of individuals in the subcritical state decreases with $\rho_0$.
At the crossover or switching point $\rho_0^s$, as indicated by the vertical
yellow dash dotted line in Fig.~\ref{fig3}, the behavior of $R(\infty)$ versus
$\lambda$ changes from being discontinuous to continuous. The crossover
point can be calculated analytically by solving
Eqs.~(\ref{stady_theta})-(\ref{First_Order}) and (\ref{First_Order_Condition}).
We also find that $\lambda_c^I$ decreases with $\rho_0$, since a large
value of $\rho_0$ enhances the probability of individuals' exposure to the
 information. In short, $R(\infty)$ versus the parameter plane
$(\rho_0,\lambda)$ shows a cusp catastrophe (i.e., the crossover phenomenon)~\cite{Strogatz2005}.
Regardless of the size of the initial seeds, there
is a good agreement between numerically calculated and theoretically
predicted behaviors of $R(\infty)$.

\subsection{Effects of topological parameters}
\label{subsec:topological_parameters}

\begin{figure}
\begin{center}
\epsfig{file=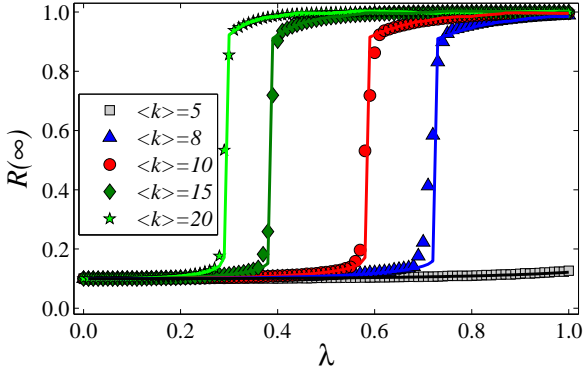,width=1\linewidth}
\caption{(Color online) {\bf Effect of mean network degree on social contagion
dynamics.} For ER random networks, $R(\infty)$ versus the transmission
probability $\lambda$ for mean degree $\langle k\rangle=5$ (gray squares),
$\langle k\rangle=8$ (blue up triangles), $\langle k\rangle=10$ (red circles),
$\langle k\rangle=15$ (dark green diamonds), and $\langle k\rangle=20$
(light green stars). Other parameters are $\rho_0=0.1$, $\gamma=1.0$ and
$T=3$. The lines are theoretical values of $R(\infty)$ calculated from
Eqs.~(\ref{active_a_k})-(\ref{s_t}) and (\ref{d_theta_2})-(\ref{r_T}).}
\label{fig4}
\end{center}
\end{figure}

We turn to elucidating the effect of network topological parameters on
social contagion dynamics in our spreading threshold model. In fact,
both the value of $R(\infty)$ and its pattern depend strongly on the
mean degree and degree heterogeneity of the network. To be concrete, we
first examine ER random networks with different values of the mean degree
$\langle k\rangle$, as shown in Fig.~\ref{fig4}, where we see that
$R(\infty)$ increases with $\langle k\rangle$ in general, since individuals
with larger degrees have higher probabilities to be exposed to
information from distinct neighbors, leading to a high likelihood that
they adopt the behavior as well. By the bifurcation analysis of Eq.~(\ref{stady_theta}),
we find that with the increase of $\langle k\rangle$,
the growth pattern of $R(\infty)$ changes from being continuous to being discontinuous.
For a small value of the mean degree (e.g., $\langle k\rangle=5$),
only a small fraction of the individuals adopt the behavior,
so $R(\infty)$ changes with $\lambda$ continuously. For
a relatively larger value of the mean degree (e.g., $\langle k\rangle>5$),
more individuals adopt the behavior, leading to a sudden, discontinuous increase in $R(\infty)$ with $\lambda$.
As discussed in Sec.~\ref{subsec:dynamical_parameters}, the
``explosive'' growth of $R(\infty)$ occurs whenever there is
a finite but sizable fraction of individuals in
the subcritical state, which cannot happen when the mean degree of the network is small.
We also observe that increasing the mean
degree can reduce the value of the critical point $\lambda_c^I$, due to
the corresponding increase in the number of individuals having relatively
large degrees.

\begin{figure}
\begin{center}
\epsfig{file=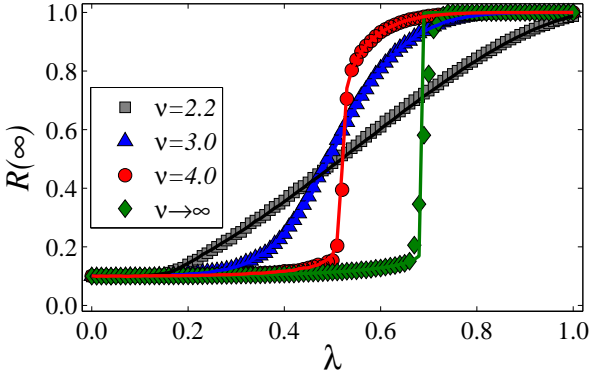,width=1\linewidth}
\caption{(Color online) {\bf Effect of network heterogeneity
on social contagion dynamics.} For scale-free networks, $R(\infty)$ versus
$\lambda$ for degree exponent $\nu=2.2$ (gray squares), $\nu=3.0$ (blue up
triangles), $\nu=4.0$ (red circles) and $\nu\rightarrow\infty$ (dark green diamonds).
The case for $\nu\rightarrow\infty$ reduces to a random
regular network with identical degree.
The lines are theoretical values of $R(\infty)$ calculated from
Eqs.~(\ref{active_a_k})-(\ref{s_t}) and (\ref{d_theta_2})-(\ref{r_T}).
Other parameters are $\rho_0=0.1$, $\gamma=1.0$ and $T=3$.}
\label{fig5}
\end{center}
\end{figure}

We next study scale-free networks. Figure \ref{fig5} shows, for $T=3$,
$R(\infty)$ versus $\lambda$ for
$\langle k\rangle=10$. The uncorrelated networks are generated with the
power-law degree distribution $P(k)\sim k^{-\nu}$ ($\nu$ being the degree
exponent) according to the procedure in Ref.~\cite{Catanzaro2005}, where
the maximum degree is set as $k_{max}\sim \sqrt{N}$. We find that increasing
the heterogeneity of network structure (by using smaller values of the degree
exponent) promotes (suppresses) $R(\infty)$ for small (large) values of
$\lambda$. This result can be qualitatively explained as follows~\cite{Wang2014A}: From Eqs.~(\ref{active_a_k})-(\ref{S_K_T}), we know that
hubs adopt the behavior with large probability. With the increase of network
heterogeneity (i.e., decreasing $\nu$), the network has a large number
of individuals with very small degrees and more individuals with large degrees.
For small values of $\lambda$, more hubs for small $\nu$ facilitate behavior spreading
as they are more likely to adopt the behavior. But for large values of $\lambda$, a large number of individuals with very small degrees have a small
probability to adopt the behavior, resulting in smaller values of $R(\infty)$.
By the bifurcation analysis of Eq.~(\ref{stady_theta}), we also observe that the
system has a critical degree exponent
$\nu^s\approx4.0$, below which $R(\infty)$ versus $\lambda$ is
continuous but above which the variation is discontinuous. That is,
as the network becomes more heterogeneous, we expect a change in the
dependence of $R(\infty)$ on $\lambda$ from being discontinuous to
continuous, since the existence of strong
degree heterogeneity can not make individuals in the subcritical state
adopt the behavior simultaneously. We also note that the critical point $\lambda_c^I$ decreases as the network
becomes more heterogeneous. Again, there is a good agreement
between theoretical and numerical results.

\section{Alternative models of social contagion dynamics}
\label{sec:other_models}

The edge-based compartmental theory developed in Sec.\ref{sec:theory} can
be applied to more general social contagion dynamics with reinforcement
effect derived from non-redundant information memory characteristic. Here, we
demonstrate the use of our theory in analyzing two alternative, somewhat
more complicated social contagion models: (1) correlated spreading
threshold model in which the adoption threshold of each individual is
correlated with his/her degree and (2) a generalized social contagion
model in which the behavior adoption probability $\pi(k,m)$
is a monotonically increasing function of $m$.

\subsection{Correlated spreading threshold model}

In reality, whether an individual adopts certain social behavior
depends on his/her personal characters such as age and habit, which
are reflected by the corresponding degree in the social network.
As a result, there is typically some correlation between an individual's
degree and his/her ability to adopt new social behaviors triggered by
crossing the adoption threshold. For simplicity, we use a recently introduced
relation~\cite{Cui2014} to account for the correlation between individual
$i$'s adoption threshold and degree $k_i$, as
\begin{equation} \label{cor_fix_th}
T_i=A_{\alpha}(\frac{k_i}{k_{max}})^{\alpha},
\end{equation}
where $k_{max}$ is the maximum degree, $A_{\alpha}$ and $\alpha$ are two
adjustable parameters. For $\alpha=0$, the adoption threshold is
uncorrelated with the degree, and all individuals in the network share the
same adoption threshold. For $\alpha > 0$, the adoption threshold is
positively correlated with the degree, i.e., individuals with larger
degrees have higher adoption thresholds, and the opposite occurs
for $\alpha < 0$.

\begin{figure}
\begin{center}
\epsfig{file=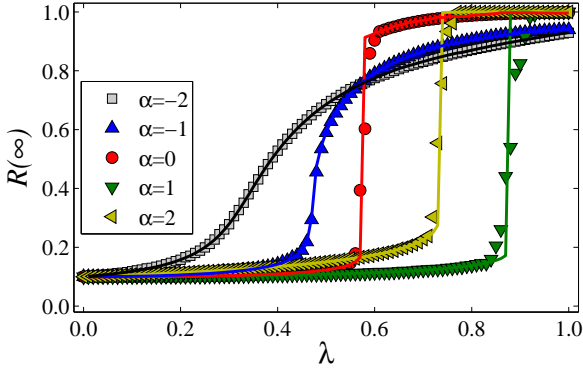,width=1\linewidth}
\caption{(Color online) {\bf Effect of degree-correlated spreading threshold
on social contagion dynamics.} For ER random networks, final adoption size
$R(\infty)$ versus the transmission probability $\lambda$ for $\alpha=-2$
(gray squares), $\alpha=-1$ (blue up triangles), $\alpha=0$ (red circles),
$\alpha=1$ (dark green diamonds), and $\alpha=2$ (yellow left triangles).
Other parameters are $\langle k\rangle=10$, $\langle T\rangle=3$, and
$\gamma=1.0$. The lines are theoretical values of $R(\infty)$ from
solutions of Eqs.~(\ref{active_a_k})-(\ref{s_t}) and
(\ref{d_theta_2})-(\ref{r_T}) with adoption threshold given by
Eq.~(\ref{A_alpha}).}
\label{figA1}
\end{center}
\end{figure}

To investigate the effects of varying $\alpha$ on social contagion
dynamics using the spreading threshold model, we set the mean adoption
threshold (somewhat arbitrarily) to be $\langle T\rangle = 3$. The sum of
the adoption threshold in the network is $T_s=\sum_{i=1}^N T_i$. For
$\alpha=0$, we have $T_s=\langle T\rangle N=A_{\alpha=0}N$. Further, we get
\begin{equation} \label{A_alpha}
A_\alpha=\frac{A_{\alpha=0}Nk_{max}^{\alpha}}{\sum_{i=1}^Nk_i^\alpha}.
\end{equation}
Evidence in terms of the quantity $R(\infty)$, which supports
our edge-based compartmental theory for varying threshold as given by
Eq.~(\ref{cor_fix_th}), is presented in Fig.~\ref{figA1}. We observe
a reasonable agreement between the theoretical predictions and simulation
results. Note that $\alpha$ affects not only the value of $R(\infty)$ but
also its dependence on $\lambda$. In particular, for $\alpha > 0$,
increasing $\alpha$ causes the critical point $\lambda_c^I$ first to
increase then to decrease. This result can be qualitatively explained
by noting that, slightly larger values of $\alpha$ (e.g, $\alpha=1$) can
cause the individuals whose degrees are near the mean degree of the network
to hold larger adoption threshold. However, much larger values of $\alpha$
(e.g., $\alpha=2$) will generate hubs with larger adoption threshold,
thereby reducing the adoption threshold for the individuals with degrees
near the mean degree. Since, in a random network, the degrees of most
individuals are close to the mean degree, this causes the non-monotonic
change in $\lambda_c^I$.

For $\alpha<0$, decreasing $\alpha$ facilitates individuals' adopting
the behavior, and the dependence of $R(\infty)$ on $\lambda$ changes from
being discontinuous to continuous by the bifurcation analysis of
Eq.~(\ref{stady_theta}). Decreasing $\alpha$ makes individuals
with small (large) degrees to hold larger (smaller) adoption thresholds
than the case of $\alpha=0$. As a result, the values of $R(\infty)$
are smaller than those for $\alpha=0$ in the large $\lambda$ regime.
Since individuals with small degrees have relatively large adoption
threshold, they have more difficulty in adopting the social behavior,
further decreasing the number of individuals in the subcritical state
and making the discontinuous behavior in $R(\infty)$ to disappear.

\subsection{A generalized social contagion model} \label{other_model}
\begin{figure}
\begin{center}
\epsfig{file=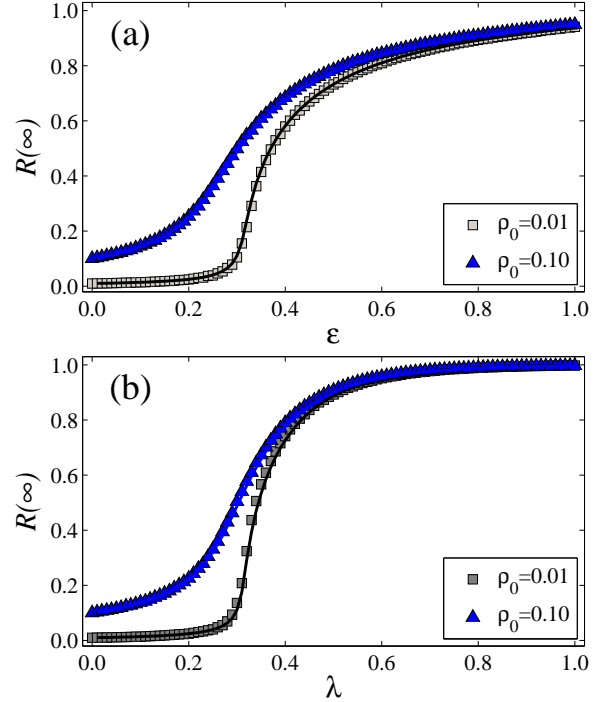,width=1\linewidth}
\caption{(Color online) {\bf Results from a generalized social contagion
model.} For ER networks, $R(\infty)$ versus the unit adoption probability
$\epsilon$ for $\lambda=0.3$ (a) and the transmission probability $\lambda$
for $\epsilon=0.3$ (b). Two values of $\rho_0$ are used: $\rho_0=0.01$ (gray
squares) and $\rho_0=0.10$ (blue up triangles). Additional parameters are
$\gamma=1.0$ and $\langle k\rangle=10$. In both panels, the lines represent
the theoretical values of $R(\infty)$ obtained from solutions of
Eqs.~(\ref{active_a_k})-(\ref{s_t}) and (\ref{d_theta_2})-(\ref{r_T})
with $\psi(t)$ given by (\ref{psi_App}).}
\label{figA2}
\end{center}
\end{figure}

Recently, Centola performed an interesting experiment of the health behavior
spreading in an online social network, and found that the behavior adoption
probability is a monotonically increasing function of $m$~\cite{Centola2010},
but not the trivial case of Heaviside step function in the spreading threshold model
and Refs.~\cite{Dodds2004,Dodds2005}. Therefore, we assume that a susceptible individual adopts the behavior with probability
\begin{equation} \label{pi_App}
\pi(k,m)=1-(1-\epsilon)^m,
\end{equation}
where $m$ is the accumulated times that the individual has been exposed
to different sources, i.e., he/she has received the  information
$m$ times from the distinct neighbors, and $\epsilon$ is the unit adoption
probability. We can also use the edge-based compartmental theory to analyze
the dynamical process of this model by substituting Eq. (\ref{pi_App})
into various equations that give the solutions of e.g., $R(\infty)$. In
particular, we rewrite Eqs.~(\ref{S_K_T}) and (\ref{neighbour_S}) as
\begin{equation} \label{S_K_T_App}
s(k,t)=\sum_{m=0}^{k}\phi_m(k,t)(1-\epsilon)^{\sum_{j=1}^m j}
\end{equation}
and
\begin{equation} \label{neighbour_S_App}
\Theta(k^{\prime},\theta(t))=\sum_{m=0}^{k^{\prime}-1}\tau_m(k^{\prime},\theta(t))
(1-\epsilon)^{\sum_{j=0}^m j},
\end{equation}
respectively, whereas Eq.~(\ref{rho_t}) has the same form as
Eq.~(\ref{rho_t_SPE}). The different aspect is that we need to replace
Eq.~(\ref{psi}) with
\begin{equation} \label{psi_App}
\begin{split}
\psi(t) &=(1-\rho_0)\sum_{k=0}^{\infty}P(k)\sum_{m=0}^{k}\binom{k}{m}(1-\epsilon)^{\sum_{i=0}^m i}\\
&\times[(k-m)\theta(t)^{k-m-1}[1-\theta(t)]^m\\
&-m\theta(t)^{k-m}[1-\theta(t)]^{m-1}].
\end{split}
\end{equation}
Substituting Eqs.~(\ref{S_K_T_App})-(\ref{psi_App}) into the corresponding
equations, we can obtain a theoretical understanding of the dynamical
evolution of the generalized social contagion model. We observe that
$R(\infty)$ varies with $\lambda$ continuously by the bifurcation analysis
of Eq.~(\ref{stady_theta}). The theoretical values
of $R(\infty)$ so predicted agree well with the simulated results,
as shown in Fig.~\ref{figA2}.

\section{Conclusions} \label{sec:conclusion}

In social contagion dynamics, memory of non-redundant information can
have a significant impact on the reinforcement mechanism required for
behavioral adoption. In particular,
the non-redundant information memory has two features:
(1) repetitive information transmission on every edge is forbidden,
(2) every individual can remember the cumulative pieces of non-redundant information.
Social reinforcement incorporating the memory characteristic
is essential to describing and understanding social contagions in the
real world. In this paper, we first proposed a general social contagion
model with reinforcement derived from this memory characteristic.
Mathematically, a model based on such
characteristic is necessarily non-Markovian. Previous works pointed out
the difficulty to develop an accurate theoretical framework to analyze
social contagion dynamics with only memory effect~\cite{Dodds2004},
let alone models with non-redundant information memory characteristic.
To meet this challenge, in this paper we developed a
unified edge-based compartmental theory to analyze social contagion
dynamics with non-redundant information memory characteristic. The
validity of our theory is established by testing it using different social
contagion models of varying complexity, different model networks.

Through a detailed study of a comparatively simple model, the spreading
threshold model, the effects of non-redundant information
memory characteristic on the social contagion
dynamics can be quantified by the final adoption size $R(\infty)$ and its
dependence on key parameters such as $\lambda$. Especially, decreasing the
adoption threshold, increasing the initial seed size or increasing the mean
degree of the network can facilitate adoption of social behaviors at the
individual level, making the system less resilient to social contagions.
The effect of structural heterogeneity on $R(\infty)$ turns out to be
more complex in that, while making the network more heterogeneous can
promote the spreading process, it impedes spreading for relatively
large values of $\lambda$. A striking phenomenon is that $R(\infty)$ as
a function of $\lambda$ can exhibit two characteristically
different types of patterns: continuous variation or sudden, discontinuous
changes, and a transition between the two patterns can be induced by
adjusting parameters such as individuals' adoption threshold, the
initial seed size or the structural heterogeneity of the network.
For example, in order to change the dependence of $R(\infty)$ on $\lambda$
from being discontinuous to continuous, we can decrease the individuals'
adoption threshold, increase the initial seed size or make the network
more heterogeneous. We also find that the discontinuous pattern
disappears when there is negative correlation between individual's adoption
threshold and his/her degree. The above crossover phenomena
can be understood through the bifurcation analysis
in theory, and also justified by analyzing the subcritical individuals in simulations.

To study social contagion dynamics in human populations is an extremely
challenging problem with broad implications and interest. Our main
contribution is a treatment of the non-redundant information memory characteristic
that is intrinsic to real world dynamics of social contagions. Our
unified edge-based compartmental theory gives reasonable understanding
of the roles of the memory characteristic in shaping the
spreading dynamics, which can be applied to analyzing
different dynamical processes such as information diffusion on computer
networks. However, many challenges remain, such as incorporation of
correlations between local structures (e.g., communities and motifs) into
social reinforcement effect at the individual level, the impacts of
redundant versus non-redundant information transmission, and
further development of analytic methods to treat non-Markovian social
contagion model on more realistic networks such as
clustered~\cite{Serrano2006,Newman2009},
multiplex~\cite{Boccaletti2014,Wang2014B,Salehi2014,Kivela2014,Lee2015},
and temporal networks~\cite{Holme2012,Barrat2013}).

\acknowledgments

This work was partially supported by the National Natural Science
Foundation of China under Grants No.~11105025, No.~61473001 and No.~91324002,
the Program of Outstanding Ph. D. Candidate in Academic Research by
UESTC under Grand No.~YXBSZC20131065, and Open Foundation of State key
Laboratory of Networking and Switching Technology (Beijing University
of Posts and Telecommunications) (SKLNST-2013-1-18). YCL was supported
by ARO under Grant No.~W911NF-14-1-0504.


\begin{thebibliography}{100}
\bibitem{Castellano2009}
C. Castellano, S. Fortunato, and S. Fortunato,
Rev. Mod. Phys. \textbf{81}, 0034 (2009).

\bibitem{Young2011}
H. P. Young, Proc. Natl Acad. Sci. USA \textbf{108}, 21285 (2011).

\bibitem{Centola2011}
D. Centola, Science \textbf{334}, 1269 (2011).

\bibitem{Banerjee2013}
A. Banerjee, A. G. Chandrasekhar, E. Duflo, and M. O. Jackson,
Science \textbf{341}, 363 (2013).

\bibitem{Centola2010}
D. Centola, Science, \textbf{329}, 1194 (2010).

\bibitem{Dodds2004}
P. S. Dodds and D. J. Watts,
Phys. Rev. Lett. \textbf{92}, 218701 (2004).

\bibitem{Dodds2005}
P. S. Dodds and D. J. Watts,
J Thor. Biol. \textbf{232}, 587 (2005).

\bibitem{Weiss2014}
C. H. Weiss, J. Poncela-Casasnovas, J. I. Glaser, A. R. Pah,
S. D. Persell, D. W. Baker, R. G. Wunderink, and L. A. NunesAmaral,
Phys. Rev. X \textbf{4}, 041008 (2014)

\bibitem{Centola2007}
D. Centola and M. Macy, Am. J. Sociol. \textbf{113}, 702 (2007).

\bibitem{Granovetter1973}
M. Granovetter,  Am. J. Sociol. \textbf{78}, 1360 (1973).

\bibitem{Watts2002}
D. J. Watts, Proc. Natl. Acad. Sci. USA \textbf{99}, 5766 (2002).

\bibitem{Newman2002}
M. E. J. Newman, Phys. Rev. E \textbf{66}, 016128 (2002).

\bibitem{Pastor-Satorras2001}
R. Pastor-Satorras and A. Vespignani,
Phys. Rev. Lett. \textbf{86}, 3200 (2001).

\bibitem{Boguna2013}
M. Bogu\~{n}\'{a}, C. Castellano, and R. Pastor-Satorras,
Phys. Rev. Lett. \textbf{111}, 068701 (2013).

\bibitem{Castellano2010}
C. Castellano and R. Pastor-Satorras,
Phys. Rev. Lett. \textbf{105}, 218701 (2010).

\bibitem{Gleeson2007}
J. P. Gleeson and D. J. Cahalane,
Phys. Rev. E \textbf{75}, 056103 (2007).

\bibitem{Whitney2010}
D. E. Whitney,  Phys. Rev. E \textbf{82}, 066110 (2010).

\bibitem{Gleeson2008}
J. P. Gleeson, Phys. Rev. E \textbf{77}, 046117 (2008).

\bibitem{Nematzadeh2014}
A. Nematzadeh, E. Ferrara, A. Flammini, and Y.-Y. Ahn,
Phys. Rev. Lett. \textbf{113}, 088701 (2014).

\bibitem{Lee2014}
K.-M. Lee, C. D. Brummitt, and K.-I. Goh, Phys. Rev. E \textbf{90},
062816 (2014).

\bibitem{Brummitt2012}
C. D. Brummitt, K.-M. Lee, and K.-I. Goh,
Phys. Rev. E \textbf{85}, 045102(R) (2012).

\bibitem{Yagan2013}
O. Ya\v{g}an and V. Gligor,
Phys. Rev. E \textbf{86}, 036103 (2012).

\bibitem{Takaguchi2013}
T. Takaguchi, N. Masuda, and P. Holme,
PLoS ONE \textbf{8}, e68629 (2013).

\bibitem{Karimi2013}
F. Karimi and P. Holme, Physica A \textbf{392}, 3476 (2013).

\bibitem{Chung2014}
K. Chung, Y. Baek, D. Kim, M. Ha, and H. Jeong,
Phys. Rev. E \textbf{89}, 052811 (2014).

\bibitem{Aral2012}
S. Aral and D. Walker,  Science \textbf{337}, 337 (2012).

\bibitem{Lv2011}
L. L\"{u}, D.-B. Chen, and T. Zhou, New. J. Phys.  \textbf{13}, 123005 (2011).

\bibitem{Ugander2012}
J. Ugander, L. Backstrom, C. Marlow, and J. Kleinberg,
Proc. Natl. Acad. Sci. USA \textbf{109}, 5962 (2012).

\bibitem{Mieghem2013}
P. V. Mieghem and R. van de Bovenkamp,
Phys. Rev. Lett. \textbf{110}, 108701 (2013).

\bibitem{Cator2013}
E. Cator, R. van de Bovenkamp, and P. V. Mieghem,  Phys. Rev. E
\textbf{87}, 062816 (2013).

\bibitem{Pastor-Satorras2014}
R. Pastor-Satorras, C. Castellano, P. V. Mieghem, and A. Vespignani,
arXiv:1408.2701v1 (2014).

\bibitem{Catanzaro2005}
M. Catanzaro, M. Bogu\~{n}\'{a}, and R. Pastor-Satorras, Phys. Rev. E \textbf{71}, 027103 (2005).


\bibitem{Francisco2011}
F. J. P\'{e}rez-Reche, J. J. Ludlam, S. N. Taraskin, and
C. A. Gilligan, Phys. Rev. Lett. \textbf{106}, 218701 (2011).

\bibitem{Karsai2014}
M. Karsai, G. I\H{n}iguez, K. Kaski,  and J. Kert\'{e}sz, J. R. Soc. Interface
\textbf{11}, 101 (2014).

\bibitem{Dodds2013}
P. S. Dodds,  K. D. Harris, and C. M. Danforth,
Phys. Rev. Lett. \textbf{110}, 158701 (2013).

\bibitem{Harris2013}
K. D. Harris, C. M. Danforth, and P. S. Dodds.
Phys. Rev. E  \textbf{88}, 022816 (2013).

\bibitem{Baxter2010}
G. J. Baxter, S. N. Dorogovtsev, A. V. Goltsev, and J. F. F. Mendes,
Phys. Rev. E \textbf{82}, 011103 (2010).

\bibitem{Zhang1989}
Y. C. Zhang, Phys. Rev. Lett. \textbf{63}, 470 (1989).

\bibitem{Miller2011}
J. C. Miller, A. C. Slim, and E. M. Volz, J. R. Soc. Interface. \textbf{10}, 1098 (2011).

\bibitem{Miller2013}
J. C. Miller and E. M. Volz, PLoS ONE, \textbf{8}(8), e69162 (2013).

\bibitem{Valdez2012}
L. D. Valdez, P. A. Macri, and L. A. Braunstein,
PLoS ONE, \textbf{7}: e44188 (2014).

\bibitem{Wang2014A}
W. Wang, M. Tang, H.-F. Zhang, H. Gao, Y. Do, and Z.-H. Liu,
Phys. Rev. E \textbf{90}, 042803 (2014).

\bibitem{Karrer2010}
B. Karrer and M. E. J. Newman, Phys. Rev. E \textbf{82}, 016101 (2010).

\bibitem{Strogatz2005}
S. H. Strogatz, \emph{Nonlinear dynamics and chaos: with applications to
physics, biology, chemistry and engineering} (Westview, Boulder, CO, 1994).

\bibitem{Erdos1959}
P. Erd\H{o}s and R\'{e}nyi, Publ. Math. \textbf{6}, 290 (1959).

\bibitem{Parshani2011}
R. Parshani, S. V. Buldyrev, and S. Havlin, Proc. Natl. Acad.
Sci. USA \textbf{108}, 1007 (2011).

\bibitem{Liu2012}
R.-R. Liu, W.-X. Wang, Y.-C. Lai, and B.-H. Wang, Phys. Rev. E \textbf{85}, 026110 (2012).

\bibitem{Achlioptas2010}
D. Achlioptas, R. M. D¡¯Souza, and J. Spencer, Science, \textbf{323}, 1453 (2009).

\bibitem{Dorogovtsev2006}
S. N. Dorogovtsev, A. V. Goltsev, and J. F. F. Mendes,
Phys. Rev. Lett. \textbf{96}, 040601 (2006).

\bibitem{Gomez-Gardenes2011}
J. G\'{o}mez-Garde\~{n}es, S. G\'{o}mez, A. Arenas, and Y. Moreno,
Phys. Rev. Lett. \textbf{106}, 128701 (2011).

\bibitem{Cui2014}
A.-X. Cui, W. Wang, M. Tang, Y. Fu, X. Liang,
Chaos \textbf{24}, 033113 (2014).


\bibitem{Serrano2006}
M. \'{A}. Serrano and M. Bogu\~{n}\'{a},
Phys. Rev. Lett. \textbf{97}, 088701 (2006).

\bibitem{Newman2009}
M. E. J. Newman, Phys. Rev. Lett. \textbf{103}, 058701 (2009).

\bibitem{Boccaletti2014}
S. Boccaletti, G. Bianconi, R. Criado, C. I. del Genio, J. G\'{o}mez-Garde\~{n}es, M. Romance,
I. Sendi\~{n}a-Nadal, Z. Wang, and M. Zanin,
Phys. Rep. \textbf{10}, 1016 (2014).

\bibitem{Wang2014B}
W. Wang, M. Tang, H. Yang, Y. Do, Y.-C. Lai, and G.W. Lee,
Sci. Rep. \textbf{4}, 5097 (2014).

\bibitem{Salehi2014}
M. Salehi, R. Sharma, M. Marzolla, M. Magnani, P. Siyari, and D. Montesi,
arXiv:1405.4329 (2014).

\bibitem{Lee2015}
K.-M. Lee, B. Mina, and K.-I. Goh, Eur. Phys. J. B \textbf{88}, 48 (2015).

\bibitem{Kivela2014}
M. Kivel\"{a}, A. Arenas, M. Barthelemy, J. P. Gleeson, Y. Moreno, and
M. A. Porter, J. Complex Netw. \textbf{2}, 203 (2014).

\bibitem{Holme2012}
P. Holme and J. Saram\"{a}ki,  Phys. Rep. \textbf{519}, 97 (2012).

\bibitem{Barrat2013}
A. Barrat, B. Fernandez, K. K. Lin, and L.-S. Young,
Phys. Rev. Lett. \textbf{110}, 158702 (2013).



\end{thebibliography}

\end{document}